\documentclass[prr,twocolumn,superscriptaddress,nofootinbib,10pt]{revtex4-2}
\usepackage{graphicx}% Include figure files
\usepackage{bm}% bold math
\usepackage{amssymb}
\usepackage{color}
\usepackage{amsmath}
\usepackage{mathrsfs} 
\usepackage{amstext}
\usepackage[polish,english]{babel}
\usepackage{latexsym}
\usepackage{array}             % Tabular
\usepackage{multirow}         % Multicolumn Tables
\usepackage[usenames,dvipsnames]{xcolor}
\usepackage[colorlinks=true,citecolor=Blue,linkcolor=RubineRed,urlcolor=Blue]{hyperref}
\usepackage{tocloft}
\usepackage{multibib}
\usepackage{listings}
\def\la{\langle}
\def\ra{\rangle}

\newcommand{\beq}{\begin{equation}}
\newcommand{\eeq}{\end{equation}}
\newcommand{\beqa}{\begin{eqnarray}}
\newcommand{\eeqa}{\end{eqnarray}}

\begin{document}
\title{Benchmarking quantum annealing dynamics: The spin-vector Langevin model}

\author{David Subires}
\affiliation{Donostia International Physics Center, E-20018 San Sebasti\'an, Spain}
\author{Fernando J. G\'omez-Ruiz\href{https://orcid.org/0000-0002-1855-0671}{\includegraphics[scale=0.45]{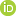}}}
\affiliation{Instituto de F\'isica Fundamental IFF-CSIC, Calle Serrano 113b, Madrid 28006, Spain}
\affiliation{Donostia International Physics Center, E-20018 San Sebasti\'an, Spain}
\author{Antonia Ruiz-Garc\'ia}
\affiliation{Instituto Universitario de Estudios Avanzados (IUdEA) and Departamento de F\'isica, Universidad de La Laguna, La Laguna 38203 Spain}
\author{Daniel Alonso\href{https://orcid.org/0000-0002-9471-2783}{\includegraphics[scale=0.45]{orcid}}}
\affiliation{Instituto Universitario de Estudios Avanzados (IUdEA) and Departamento de F\'isica, Universidad de La Laguna, La Laguna 38203 Spain}
\author{Adolfo del Campo}
\email{adolfo.delcampo@uni.lu}
\affiliation{Department of Physics and Materials Science, University of Luxembourg, L-1511 Luxembourg City,  Luxembourg}
\affiliation{Donostia International Physics Center, E-20018 San Sebasti\'an, Spain}

\begin{abstract}
The classical spin-vector Monte Carlo (SVMC) model is a reference benchmark for the performance of a
quantum annealer. Yet, as a Monte Carlo method, SVMC is unsuited for an accurate description of the annealing
dynamics in real-time.We introduce the spin-vector Langevin (SVL) model as an alternative benchmark in which
the time evolution is described by Langevin dynamics. The SVL model is shown to provide a more stringent test
than the SVMC model for the identification of quantum signatures in the performance of quantum annealing
devices, as we illustrate by describing the Kibble-Zurek scaling associated with the dynamics of symmetry
breaking in the transverse field Ising model, recently probed using D-Wave machines. Specifically, we show that
D-Wave data are reproduced by the SVL model.\\
\\
DOI: \href{https://link.aps.org/doi/10.1103/PhysRevResearch.4.023104}{10.1103/PhysRevResearch.4.023104}
\end{abstract}
\maketitle

\section{Introduction}
Adiabatic quantum computing  provides an approach to solve optimization problems by utilizing the quantum dynamics generated by a time-dependent Hamiltonian. The latter is chosen to interpolate between an initial Hamiltonian with a ground-state that can be easily prepared (e.g. a paramagnet) and a final noncommuting Hamiltonian, the ground state of which encodes the solution to the optimization problem~\cite{kadowaki_quantum_1998,Brooke99,Farhi01,Santoro02}.  While the success of the computation relies intuitively on fulfilling adiabaticity during the quantum annealing dynamics, this condition is generally not fulfilled in real devices~\cite{AlbashLidar18}. A relevant example is that of D-Wave machines utilizing time-dependent Hamiltonians of the Ising type with a transverse field \cite{Harris10,DWave-entanglement,Johnson2011,Boixo13,Boixo14}.

State-of-the-art quantum annealers are an example of noisy intermediate-scale quantum (NISQ) devices~\cite{Preskill18} in which 
various sources of noise can give rise to decoherence~\cite{Amin09,Albash15}. The latter results from the buildup of quantum correlations between the degrees of freedom described by the interpolating Hamiltonian and the surrounding environment, which is generally inaccessible and hard to characterize. Using the formalism of open quantum systems~\cite{BreuerBook}, a quantum master equation prescribes in this scenario the evolution of the state of the system, which is encoded in a density matrix.

Decoherence is broadly acknowledged as being responsible for the emergence of classical behavior in quantum systems~\cite{Zurek03}. As such, it hampers the potential of quantum computers to exhibit a quantum advantage over their classical counterpart. Benchmarking the performance of quantum annealers against classical models has thus become a central goal. Efforts to this end consider models of interacting classical rotors~\cite{Smolin14,Vazirani14a,Vazirani14,Albash15a,Albash15b,AlbashHen15,Bando20,AlbashMarshall21}. The spin-vector Monte Carlo (SVMC) model constitutes a paradigmatic reference in which the dynamics is implemented via Monte Carlo steps. Due to the difficulty to relate Monte Carlo steps to real-time evolution, the SVMC model is limited as a benchmark for the annealing dynamics. Circumventing this limitation requires a description of the annealing dynamics in continuous time. In this context, dissipative Landau-Lifshitz-Gilbert equations resembling those used in magnetism~\cite{Landau2,Landau1,Gilbert} have been put forward~\cite{Wang13,Crowley14,Crowley16}. Further progress has been achieved considering the evolution of spin-coherent states~\cite{DenchevPRX}.

Decoherence has also motivated the benchmarking  of quantum simulators and annealers with models of open quantum systems~\cite{Lanting11,Albash12,Boixo13,Albash15a,Albash15,AlbashHen15,Amin15,Boixo16,Passarelli20,Bando20,AlbashMarshall21}. Embedding the  problem Hamiltonian in a  harmonic environment, the spin-boson model has been utilized to assess the performance of D-Wave machines. Once the evolution is no longer considered to be unitary, the set of candidate quantum channels that can account for it is rich. Yet, dissipative classical systems are also rich, and it appears that efforts to accommodate dissipative effects in a quantum description of annealing devices have not been accompanied by comparable efforts in the classical domain, which may provide stringent tests for the identification of intrinsically quantum features in the annealing dynamics.

Classical Langevin dynamics provides a natural setting to describe evolution in real-time and accommodates the interplay between dissipation and thermal fluctuations~\cite{Allen2017}. 
In this work, we introduce the spin-vector model evolving under Langevin dynamics, which we shall refer to as the spin-vector Langevin (SVL) model. We use it to describe the dynamics across the phase transition in the transverse-field Ising model, recently used to probe the dynamics in quantum annealers in D-Wave devices~\cite{Gardas2018,Weinberg20,Bando20}. We characterize the scaling of the mean number of kinks with the annealing time as well as the kink number distribution, and we conclude that the performance of the D-Wave machines is reproduced by the SVL model.

\section{The SVL Model}
Consider an ensemble of $N$ qubits distributed on a graph $G(E,V)$ with edge and vertex sets denoted by $E$ and $V$, respectively.
Quantum annealing is based on the dynamics generated by the time-dependent Hamiltonian
\beqa
\label{QAH}
H(t)=A(t)H_0+B(t)H_P,
\eeqa
where the initial Hamiltonian reads $H_0=-\sum_{i\in V}\sigma_i^x$ and the problem Hamiltonian is of the Ising type
\beqa
H_P=-\sum_{(i,j)\in E}J_{ij}\sigma_i^z\sigma_j^z-\sum_{i\in V}g_i\sigma_i^z,
\eeqa
although more general forms can be considered.
Here,  $\sigma_{i}^z$ is the Pauli operators acting on vertex $i$. The constant $g_i$ plays the role of a local magnetic field,
while the spin-spin couplings $J_{ij}$ can favor ferromagnetic ($J_{ij}>0$) or antiferromagnetic order ($J_{ij}<0$).
The real functions $A(t)$ and $B(t)$ satisfy the boundary conditions $A(0)=1$, $B(0)=0$, $A(t_a)=0$, $B(t_a)=1$, where $t_a$ is the annealing time.
The goal is to find the ground state of the problem Hamiltonian upon completion of the protocol at time $t=t_a$. 
%This process is assisted by the noncommuting term $H_0$ favoring tunneling. 
%Finite temperature effects and sources of decoherence are to be expected in actual devices.
The spin-vector model is a classical annealing Hamiltonian obtained by replacing Pauli operators by real functions of a continuous angle $\theta$, i.e.,
$\sigma_i^z\rightarrow \sin\theta_i $ and $ \sigma_i^x\rightarrow\cos\theta_i$. Thus, each vertex is associated with a classical planar rotor.
The SVMC model unravels the classical dynamics of the planar rotors via Monte Carlo  steps. There is no unique recipe to relate these steps to the flow of continuous-time in real dynamics.

As an improved benchmark to assess the classicality of the performance of a quantum annealer, we propose the SVL model, in which Monte Carlo steps are replaced by Langevin dynamics.
The configuration of the system is thus specified by the set of angles ${\bm \theta}=(\theta_1,\dots,\theta_N)$
 and the dynamics is described by the stochastic coupled equations of motion
\beqa
m_i\ddot{\theta_i}+\gamma\dot{\theta}_i+\frac{\partial H({\bm \theta})}{\partial\theta_i}+\xi_i(t)=0,\quad i=1,\dots,N,
\label{SVLEq}
\eeqa
where explicit computation yields 
\beqa
\frac{\partial H({\bm \theta})}{\partial\theta_i}=-B\sum_{j\in V}^N J_{ij}\cos\theta_i \sin\theta_{j}-Bg_i\cos\theta_i+A\sin\theta_i.\nonumber\\
\eeqa
Here, $\xi_i(t)$ is an iid Gaussian real process acting on the $i$-th rotor, $m_i$ is an effective mass that provides its inertia, and $\gamma$ is the damping constant. We consider the fluctuation-dissipation relation
$ \la\xi_i(t)\xi_j(t')\ra=2\gamma k_BT\delta_{ij}\delta(t-t')$, where $T$ is the temperature of the thermal reservoir \cite{Kubo66,SM}. In what follows, we consider the case $m_i=m$ for all rotors and work in units with $k_B=1$. 
Non-Markovian variants can be accommodated for by replacing the $\delta(t-t')$ function by a memory function $\chi(t-t')$. 
The numerical integration of the SVL equations of motion is detailed in Appendixes \ref{AppA} and \ref{AppB}, and implemented by the  FORTRAN code {\sf SVLdynamics.f}  available as supplemental material.

\section{SVL and the Transverse Field Ising Model (TFIM)}
We next focus on the one-dimensional TFIM as a case study, used to benchmark the annealing dynamics in D-Wave systems \cite{Gardas2018,Bando20}.
%This corresponds to choosing the graph $G(E,V)$ as a path graph $P_N$ under open boundary conditions, with an edge set restricted to nearest neighbors.
The quantum TFIM Hamiltonian reads
\beqa
H(t)=-J\sum_{i=1}^{N-1}\sigma_i^z\sigma_{i+1}^z-h\sum_{i=1}^N\sigma_i^x,
\eeqa
with homogenous ferromagnetic coupling $J>0$ and magnetic field $h$.
In quantum annealing, this corresponds to a toy-model scenario in which the problem Hamiltonian is a one-dimensional homogeneous ferromagnet $H_P=-J\sum_{i=1}^{N-1}\sigma_i^z\sigma_{i+1}^z$.
The model is exactly solvable under periodic boundary conditions and exhibits a quantum phase transition signaled by the closing of the energy gap between the ground state and the first excited state \cite{LSM61}.
As a result, the correlation length $\xi$ and the relaxation time $\tau$ exhibit the characteristic power-law divergence of critical systems, 
$\xi=\xi_0/|\epsilon|^\nu$ and $ \tau=\tau_0/|\epsilon|^{z\nu}$, where $\xi_0$ and $\tau_0$ are microscopic constants and $\epsilon\propto h-h^*$ is the distance to the critical point, which equals $h^*=J$ for large $N$. 
The location of the critical point is the same in the classical and the quantum TFIM.
For the quantum TFIM in isolation, the correlation length critical exponent $\nu=1$ and the dynamic critical exponent $z=1$. 
\begin{figure*}[t]
	\centering
		\includegraphics[width=0.8\linewidth]{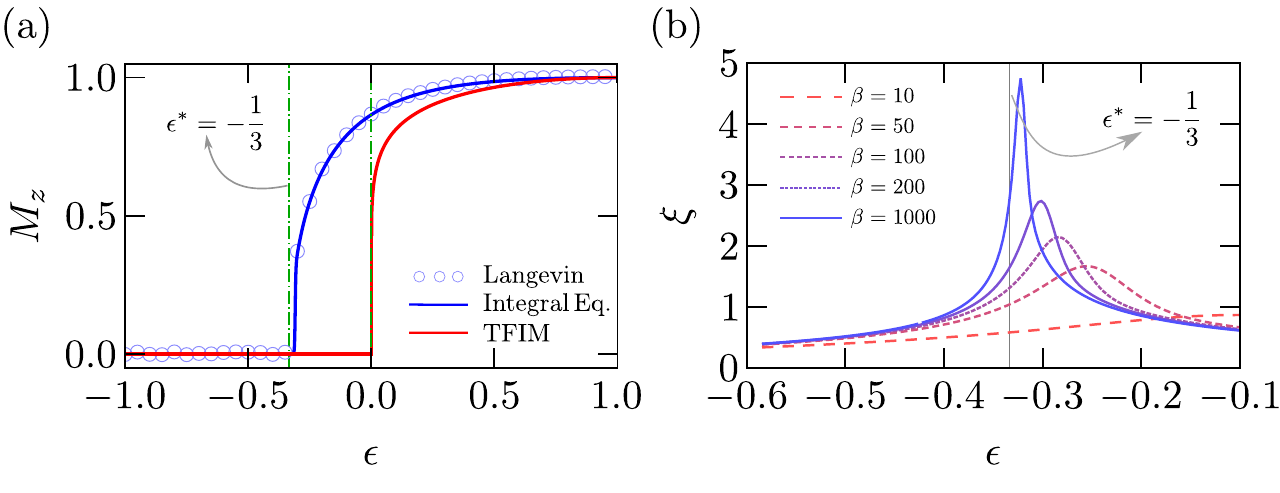}
	\caption{\label{figSVLeq}Equilibrium properties of the SVL description of the TFIM. (a) Growth of the averaged absolute value of the local longitudinal magnetization $M_z$ (order parameter) as a function of the parameter $\epsilon=J-h$ in the SV model at zero temperature. The critical value is $\epsilon^*=-1/3$, in contrast with the quantum TFIM for which $\epsilon^*=0$. The behavior is reproduced by the SVL simulations at $T=10^{-3}$. The shift in $\epsilon^*$ can be used to distinguish whether the degrees of freedom in a given simulator are described by planar rotors or binary spins.
(b)  Equilibrium correlation length $\xi$ as a function of the parameter $\epsilon$ for different values of the inverse temperature $\beta$.
}	
\end{figure*}

By contrast, in the SVL description, the Mermin-Wagner theorem precludes spontaneous symmetry breaking at finite temperature. At zero-temperature, a transfer matrix analysis shows that the critical point is located at $h^*=J+1/3$, see Appendix \ref{AppC} for a detailed derivation. This is illustrated in Fig. \ref{figSVLeq} in which the averaged absolute value of the local longitudinal magnetization $M_z=\frac{1}{N}\sum_{i=1}^N\la|\sin\theta_i|\ra$, which acts as the order parameter, is shown as  a function of $\epsilon=J-h$. This equilibrium behavior is captured by the long-time dynamics of the SVL model in the low-temperature limit. 
By monitoring the equilibrium value of $M_z$ one can thus distinguish the description in terms of planar rotors used in SV models from binary spins in a given NISQ device, such as a D-Wave machine.
The growth of $M_z$ is typical of a continuous phase transition and is accompanied by the power-law scaling of the correlation length observed as the zero-temperature limit is approached, revealing the correlation-length critical exponent $\nu=1/2$.

Further, by varying the strength of the damping constant $\gamma$, Langevin dynamics interpolates between Hamiltonian and diffusive dynamics.
To identify the value of the dynamic critical exponent $z$, we consider the linearized system, setting $\sin \theta\approx \theta$ and $\cos \theta\approx 1-\theta^2/2$, 
$H({\bm \theta})=-\sum_{i=1}^{N-1} J\theta_i \theta_{i+1}+\frac{h}{2} \sum_{i=1}^N \theta_i^2-hN$.
For the 1D homogeneous Ising chain with nearest neighbor interactions 
\beqa
m\ddot{\theta_i}+\gamma\dot{\theta}_i+h\theta_i-J(\theta_{i-1}+\theta_{i+1})+\xi_i(t)=0,\forall i.
\eeqa
In the overdamped regime, $\tau\simeq\left|\theta_{j}/\dot{\theta}_j\right|\simeq\frac{\gamma}{h-2J}$,
and thus $z=2$ and $\tau_{0}=\gamma$.
Similarly, in the underdamped regime,
$\tau\simeq\left|\theta_{j}/\ddot{\theta}_j\right|^{1/2}\simeq\left|\frac{m}{h-2J}\right|^{1/2}$
which implies $z=1$ and $\tau_0=\sqrt{m}$.
We note that these values of $z$ are consistent with mean-field values derived for the Ginzburg-Landau equation in the corresponding overdamped and underdamped regimes \cite{LagunaZurek98,delcampo10}. As we shall see, a continuous range of effective intermediate values, $z\in[1,2]$, can be spanned by varying the damping constant $\gamma$. 
For completeness, we note that the model  in \cite{Smolin14} is time-continuous,   with $\gamma=0$ and fixed noise strength. This is inconsistent with the fluctuation-dissipation theorem,  bringing the system to an infinite-temperature state, and fixes the value of $z=1$, with no freedom, as in the SVMC.

\section{Benchmarking critical dynamics via the Kibble-Zurek mechanism (KZM)}
\begin{figure}[t]\centering
\includegraphics[width=0.8\linewidth]{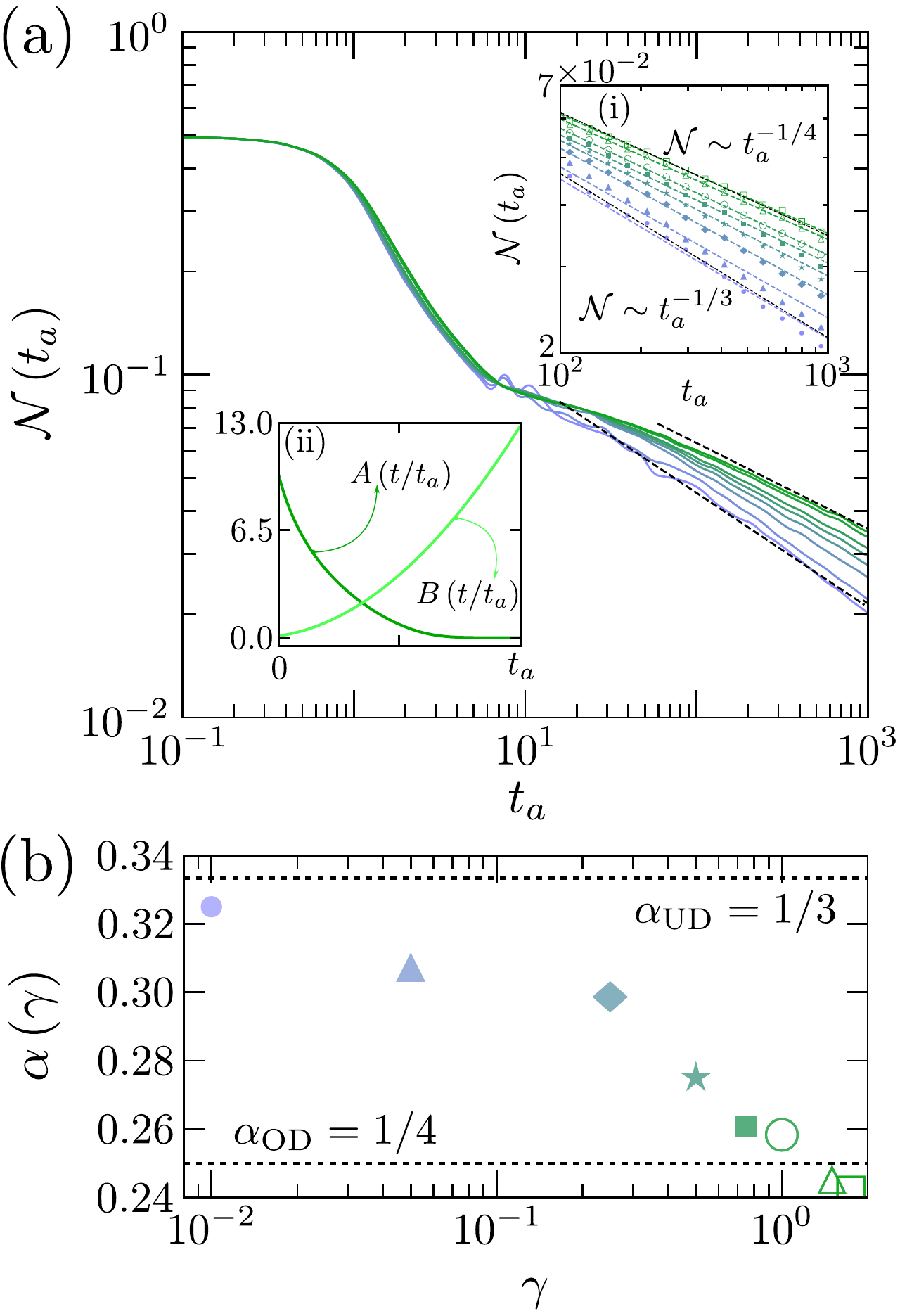}
\caption{\label{FigSVL1} (a) Scaling of the average kink density in the SVL model as a function of the annealing time $t_a$, for different values of the damping constant $\gamma\in[0.01,5]$. Data are collected from $10^4$ stochastic trajectories. The inset (i) shows in detail the scaling regimen spanning the range between the power laws $t_a^{-1/3}$ and $t_a^{-1/4}$. Inset (ii) shows the time dependence of the coefficients $A$ and $B$ during the annealing schedule. (b)  Power-law exponent $\alpha$ as a function of the damping constant $\gamma$. The dotted lines indicate the KZM prediction in the underdamped and overdamped limits.  The SVL model reproduces a broad range of KZM exponents including reported values in D-Wave devices and that of the spin-boson approach.
}
\end{figure}

According to the celebrated Kibble-Zurek mechanism (KZM)~\cite{Kibble76a,Kibble76b,Zurek96a,Zurek96c,DZ14}, the crossing of a phase transition results in topological defects. The KZM 
yields a universal power-law scaling of the density of defects as a function of the annealing time $t_a$
$\mathbb{E}(\mathcal{N})\sim t_a^{-\alpha_{\rm KZM}}$, where $\alpha_{\rm KZM}=\frac{d\nu}{1+ z\nu}$, in spatial dimensions $d$ and point-like defects. This prediction is a natural test for benchmarking the dynamics in a quantum simulator \cite{Gardas2018,Weinberg20,Bando20}. 
In the TFIM, the transition between a paramagnet and a ferromagnet 
results in the formation of $\mathbb{Z}_2$-kinks. The latter can be detected by the kink-number operator \cite{Dziarmaga05} $\mathcal{N}=\frac{1}{2}\sum_{i=1}^{N-1}\left(1-\sigma_i^z\sigma_{i+1}^z\right)$.
For the SVL, we consider its analog 
$\mathcal{N}=\frac{1}{2}\sum_{i=1}^{N-1}\left[1-{\rm sgn}(\sin\theta_i){\rm sgn}(\sin\theta_{i+1})\right]$,
where the sign function is included for proper counting, establishing a mapping from the continuous-variable description of each spin to a binary configuration.
As the values of $z$ and $\alpha$ are sensitive to the presence of dissipation \cite{LagunaZurek98,delcampo10}, they can be used to distinguish between unitary evolution and open dynamics in quantum systems \cite{Patane08}.
In the quantum TFIM in isolation, $\alpha_{\rm KZM}=1/2$ \cite{Polkovnikov05,Dziarmaga05,Zurek2005}. Coupling to a bath can result in anti-KZM behavior associated with heating \cite{Dutta16} as observed in \cite{Weinberg20}. It can also preserve KZM behavior while leading to a more subtle renormalization of the critical exponents \cite{Pankov04,Sachdev04}. Data collected in D-Wave machines for the critical dynamics of the 1D TFIM are described by $\alpha=0.20$ in the NASA machine and by $\alpha=0.34$ in the Burnaby device~\cite{Bando20}, inconsistent with the unitary evolution of the quantum 1D TFIM. The values are also inconsistent with classical models including the SVMC model~\cite{Bando20}, simulated quantum annealing~\cite{Bando21}, and Glauber dynamics~\cite{Mayo21}, and they have been explained using a spin-boson model, coupling the quantum TFIM to an Ohmic harmonic bath. In this case, the theoretical value $\alpha_{\rm KZM}=0.28$ while a broader range is found numerically at zero temperature varying the spectral function \cite{Bando20}. 
The theoretical KZM exponents in the SVL model cover the range $\alpha_{\rm KZM}\in[1/4,1/3]$ by decreasing the damping constant from the overdamped to the underdamped regime. As in the recent D-Wave tests, we report the kink density upon completion of the annealing schedule at $T=t_a$. For long annealing times, SVL numerics corroborates the KZM prediction taking into account the dependence of the dynamic critical exponent $z$ on $\gamma$, as shown in Fig. \ref{FigSVL1}. The range of scaling exponents reproduced by the SVL model thus includes the values reported for the spin-boson model and the Burnaby device. That of the NASA device (0.20) is slightly lower than that in the overdamped limit (0.25).  
 
 As a caveat, it should be taken into account that other effects can alter the KZM scaling. In either classical or quantum systems, nonlinear modulations of the transverse field \cite{Diptiman08,Barankov08}, ubiquitous in D-Wave machines, as well as spatially inhomogeneous controls \cite{DM10,delcampo10,DKZ13,Fernando19} and the presence of quench disorder \cite{Dziarmaga2006} can modify the scaling behavior. 
 These effects are negligible in our simulations, but they could be present in simulators such as D-Wave devices.
 However, the choice of the measurement time for the KZM to apply is not trivial.
 It should exceed the freeze-out time scale $\hat{t}\sim(\tau_0t_a^{z\nu})^{\frac{1}{1+z\nu}}$, although the choice of the prefactor remains open. The latter cannot be too large as other effects such as defect annihilation and coarsening can compete and even hide the KZM, given that the SVL description involves coupling to a thermal bath. These conditions are not always guaranteed when probing the state right after upon completion of the schedule, at $t_a$, as done in Fig. \ref{FigSVL1}.
 Numerical simulations for the SVL model reveal that the growth of the order parameter generally lags behind the completion of the annealing schedules for fast and moderate quenches, as shown in  Appendix \ref{AppD}. Only for slow quenches is the scaling consistent with the KZM prediction. 
\begin{figure}[t]\centering
\includegraphics[width=0.8\linewidth]{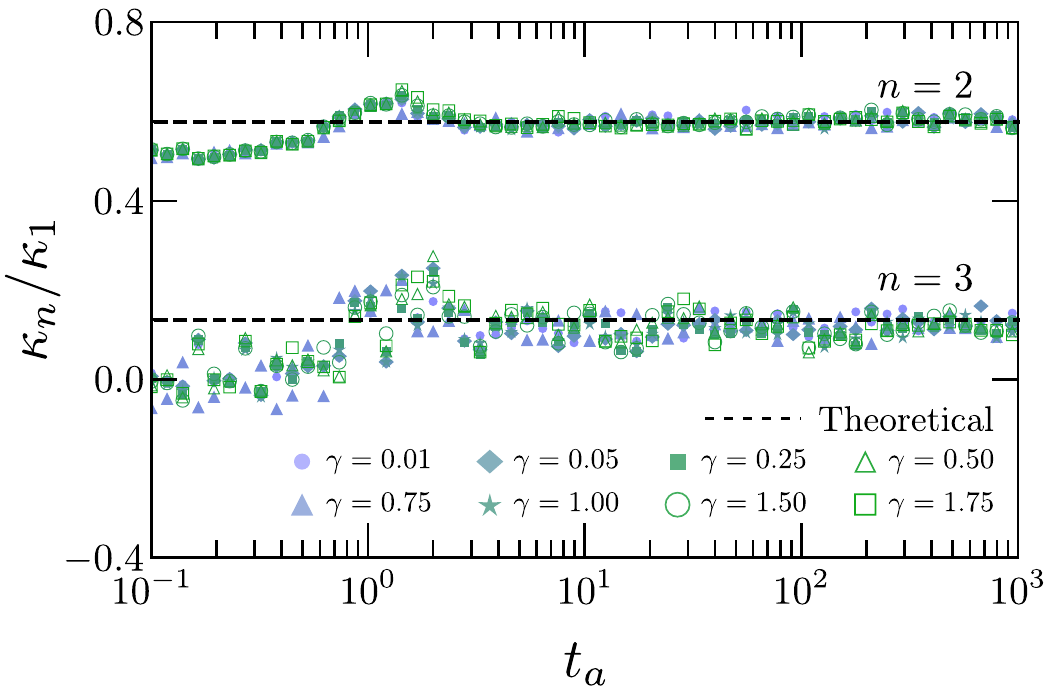}
\caption{\label{figSVLcr} The cumulant ratios $\kappa_2/\kappa_1$ and $\kappa_3/\kappa_1$ upon completion of the annealing schedule at $t=t_a$ for different values of the damping constant, as a function of the annealing time $t_a$ (using $10^4$ stochastic trajectories). The dashed lines indicate the expected theoretical value for each cumulant ratio. Cumulant ratios predicted by the SVL model reproduce values previously attributed to quantum models, in isolation or open.}
\end{figure}

\section{Kink statistics beyond KZM}
For a deeper characterization of the annealing dynamics, we consider the probability distribution to find $n$ kinks in the final nonequilibrium state prepared upon completion of the annealing protocol, 
$P(n)=\mathbb{E}[\delta(\mathcal{N}-n)]$.
In the quantum domain, for the TFQIM evolving under unitary dynamics, an exact analytical computation shows that $P(n)$ is Poisson-binomial distribution \cite{delcampo18,Cui19}. 
The Fourier transform of $P(n)$ is the  
characteristic function $\tilde{P}(\varphi)=\mathbb{E}[e^{i\varphi n}]$ and its logarithm is the cumulant generating function, which admits the expansion
$
\ln\mathbb{E}[e^{i\varphi n}]=\sum_{p=1}^\infty\frac{\kappa_p}{p!}(i\varphi)^p$, where $\kappa_p$ is the cumulant of order $p$.
The key prediction of physics beyond KZM \cite{delcampo18,GomezRuiz19b} is that $\kappa_q\propto\kappa_1$ and thus $\kappa_q/\kappa_1$ are constant and independent of the annealing time.  
For the quantum TFIM in isolation it was shown that $\kappa_2/\kappa_1=2-\sqrt{2}\approx0.578$ and $ \kappa_3/\kappa_1=4-12/\sqrt{2}+8/\sqrt{3}\approx0.134$. 
Their study can be used to rule out models of the underlying dynamics \cite{Bando20,Bando21,Mayo21}.
However, cumulant ratios can be robust to decoherence
 as shown by simulations of the spin-boson quantum model with independent oscillators being coupled to the $z$ component of each spin \cite{Bando20}. 
The cumulant ratios in the SVL are shown in Fig. \ref{figSVLcr}. While they exhibit a dependence on the annealing time for fast schedules, their value soon saturates at the theoretical prediction for the TFIM not only at long annealing times but even before the KZM scaling regime sets in, for moderate annealing times. Again, the SVL model reproduces the observed values in D-Wave. 
This is consistent with the generalized KZM \cite{GomezRuiz19b,Mayo21} according to which the dynamics sets the correlation length out of equilibrium $\hat{\xi}$, and defect formation can be described as the result of a sequence of $\sim N/\hat{\xi}$ iid discrete random variables, that yields a binomial distribution for $P(n)$.
Cumulant ratios are then set by the success probability for a kink formation, which can be expected to be weakly dependent on the underlying dynamics. Indeed, the latter is exclusively dictated by the structure of the vacuum manifold and geometric arguments when invoking the geodesic rule \cite{Kibble76a}. 

 %%%%%%%%%%%%%%%%%%%%%%%%%%%%%%%%%%%%%%%%%%%%%%%

\section{Discussion}
The SVMC method constitutes an important benchmark for the performance of a quantum annealer, in which each quantum spin is replaced by a classical planar rotor. As a Monte Carlo method is ill-suited to describe real-time dynamics, we have introduced the SVL model in which discrete Monte-Carlo updates are replaced by Langevin dynamics, which is stochastic, continuous in time, and governed by the fluctuation-dissipation theorem. 

We have shown that the SVL annealing dynamics yields a power-law scaling for the average density of topological defects as a function of the annealing time, using the 1D TFIM as a case study. At variance with Monte Carlo  methods, the power-law exponent continuously interpolates between the KZM prediction for the underdamped and overdamped regimes. 
 Remarkably, the SVL dynamics spans the power-law exponents observed in D-Wave machines (away from the fast-annealing limit) and the spin-boson approach, in the classical realm. 
Beyond the Kibble-Zurek scaling, we have analyzed the kink number statistics in which all cumulants share the same power-law scaling with the annealing time. Cumulant ratios are thus fixed and those reported in D-Wave are further reproduced by the SVL model.
We expect the SVL model to provide a test for classicality of quantum annealers and programmable simulators based on Ising spin models, as it can be extended to arbitrary graphs, inhomogeneous rotors,
and non-Markovian dynamics, e.g., accounting for the $1/f$ noise that is expected to be relevant for long annealing times \cite{DWaveInstr}. More generally, our results advance the use of Langevin methods to benchmark NISQ devices in quantum computing and quantum simulation.

{\it Note added.} After the completion of  the work, King et al.~\cite{King22} reported data collected in  D-Wave 2000Q lower noise processor for the fast annealing dynamics of the TFIM, in agreement  with the theoretical analysis for the unitary evolution of an isolated spin chain~\cite{delcampo18,Cui19}.

\acknowledgements
 It is a pleasure to thank Andrew King for discussions. We further thank Hidetoshi Nishimori for a careful reading of the manuscript. F.J.G.R acknowledges the hospitality of the University of Luxembourg during the completion of this work. 
This project has been funded by the Spanish MINECO and the European Regional Development Fund FEDER
through Grant No. FIS2017-82855-P (MINECO/FEDER,UE).
\appendix

\section{SVL dynamics }\label{AppA}

The dynamics of the rotors in contact with a Langevin thermostat can be described by the $2N$-dimensional stochastic differential equations
\begin{equation}\label{eqn2}
\begin{split}
d\theta_{i}&=\frac{p_{i}}{m}\,dt,\:\quad\text{with}\quad (i=1,\dots,N), \\
dp_{i}&=-\left(\frac{\partial H({\bm \theta})}{\,\,\partial\theta_{i}}+\frac{\gamma}{m}\,p_{i}\right)dt+\sqrt{2D}\,\,dW_{i}.  
\end{split}
\end{equation}
where $\gamma$ and $D$ are the friction and diffusion coefficients associated with the interaction with the thermal bath. According to the fluctuation-dissipation theorem \cite{Kubo66}, both coefficients are related according to 
\begin{equation}
D\,=\gamma k_B T\,,
\end{equation}
where $k_B$ is the Boltzmann constant. The term $dW_{i}$ denotes a Wiener process resulting from the Gaussian white noise force $\xi_{i}(t)$ acting on the $i$-th rotor and associated with the diffusion. These Wiener processes satisfy
\begin{equation}
\langle\xi_{i}(t)\rangle\,=\,0,
\end{equation} 
and 
\begin{equation}
\langle\xi_{i}(t)\,\xi_{j}(t^\prime)\rangle\,=\,2D\,\delta_{ij}\delta(t-t^\prime)\,,
\end{equation}
where $\langle\dots\rangle$ denotes the statistical average.

\section{Numerical integration of the stochastic SVL equations of motion}\label{AppB}
The supplemental material includes the Fortran code {\sf SVLdynamics.f} for the numerical integration of the SVL equations of motion, that we next describe.
The $2N-$dimensional stochastic differential equations (\ref{eqn2}) can be expressed in the matrix form 
\begin{equation}\label{eq}
d{\bf Y}={\bf A}({\bf Y})\,dt\,+\,{\bf B}\cdot d\bm{\Omega}_t\,,
\end{equation}
where the components of the variable vector ${\bf Y}$ have been ordered as
\begin{equation}
{\bf Y}=(\theta_{1},\dots,\theta_{N},{p}_{1},\dots{p}_{N})\,.
\end{equation} 
The components of the vector ${\bf A}$ containing the deterministic terms in the equations of motion are
\begin{eqnarray}
A_{i}=
\left\{
\begin{array}{ll} 
\frac{p_i}{m} & \quad i=1,\dots,N\,, \nonumber\\
-\left(\frac{\partial H({\bm \theta})}{\partial \theta_{i-N}}+\,\frac{\gamma}{m} p_{i-N}\right) & \quad i=N+1,\dots,2N\,.
 \end{array}
 \right.
\end{eqnarray}
The matrix ${\bf B}$ contains the diffusion coefficients $D$. In our model it is given by a diagonal matrix with the elements
\begin{eqnarray}
B_{ii}=
\left\{
\begin{array}{ll} 
0 & \quad i=1,\dots,N\, \\
\sqrt{2D} & \quad i=N+1,\dots,2N\,.
 \end{array}
 \right.
\end{eqnarray}
The vector $d\bm{\Omega}_t$ denotes the $2N-$dimensional Wiener process, with the elements
\begin{eqnarray}
d\Omega_{t,i}=
\left\{
\begin{array}{ll} 
0 & \quad i=1,\dots,N\,, \\
dW_{i-N} & \quad i=N+1,\dots,2N\,.
 \end{array}
 \right.
\end{eqnarray}
To integrate the stochastic differential equations (\ref{eq}), we consider the multi-dimensional explicit order 2.0 weak scheme \cite{Platen11}. Since in our model the matrix ${\bf B}$ does not depend explicitly on the variable vector ${\bf Y}$, such scheme becomes particularly simple. Specifically, given the variable ${\bf Y}_{I}$ at a time step $I$, its value at the following time step $I+1$ is given by
\begin{equation}
{\bf Y}_{I+1}={\bf Y}_{I}+\frac{1}{2}\left[\,{\bf A}(\bm{\Gamma}_{I})+{\bf A}({\bf Y}_{I})\,\right]\Delta_t+{\bf B}\cdot\Delta\bm{\Omega}_{I}\,,
\end{equation}
where 
\begin{equation}
\bm{\Gamma}_{I}={\bf Y}_{I}+{\bf A}({\bf Y}_{I})\Delta_{t}+{\bf B}\cdot\Delta\bm{\Omega}_{I}\,,
\end{equation} 
with $\Delta_{t}=t_{I+1}-t_{I}$ the constant time interval between two consecutive time steps, and $\bm{\Omega}_{I}$ the vector with elements
\begin{eqnarray}
\Delta\Omega_{I,i}=
\left\{
\begin{array}{ll} 
0 & \quad i=1,\dots,N\,, \\
\sqrt{\Delta_{t}}\,\,G_{I,i-N}, & \quad i=N+1,\dots,2N\,,
 \end{array}
 \right.
\end{eqnarray}
where $G_{I,n}\sim N(0,1)$ is a normally distributed random variable assigned for the $n$-rotor at time step $I$.

\section{Equilibrium properties of the SV model}\label{AppC}
The SV model associated with the TFIM describes a linear chain of planar rotors with homogenous coupling between nearest neighbors and a global magnetic field, 
\begin{align}
\label{HSVIsing}
  H({\bm \theta})=-J\sum_{i=1}^{N} \sin\theta_i \sin\theta_{i+1}-h \sum_{i=1}^N \cos \theta_i,
\end{align}
with $i\in\{1,\cdots,N\}$, $\theta_{N+1}=\theta_1$ and $J,h \ge 0$.  This corresponds to choosing the graph $G(E,V)$ in the general problem Hamiltonian as a path graph $P_N$ under open boundary conditions, with an edge set restricted to nearest neighbors.
By construction, the system has a $\mathbb Z_2$ symmetry with respect to the involution $\Pi (\theta_1,\cdots,\theta_N)=(-\theta_1,\cdots,-\theta_N)$.
To characterize the thermal equilibrium properties of the SVL model, we make use of the transfer matrix technique \cite{Mattis84}, which provides an alternative to study thermalization by long-time Langevin dynamics, and allows for a deeper analytical treatment. In our analysis we have made use of both methods, and we verify a good agreement between them. We rewrite the Hamiltonian (\ref{HSVIsing})
as
\begin{figure*}[t!]
\includegraphics[width=0.9\linewidth]{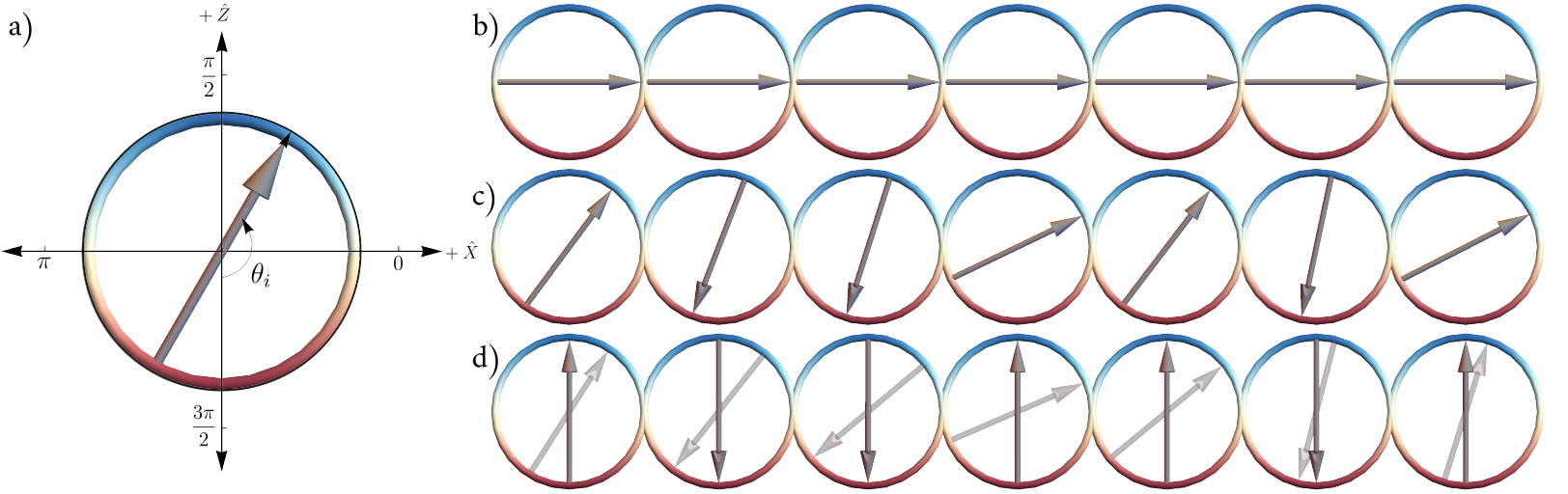}
\caption{\label{FigScheme} Schematic representation of the counting of kinks in a given configuration of planar rotors. (a) In the SVL each spin is represented by a planar rotor. (b) All rotors are aligned in the paramagnetic phase, the initial state of the annealing schedule. (c) A non-equilibrium state results upon completion of the schedule. (d) A sign function maps the angle of each rotor to a binary variable facilitating the counting of kinks. }
\end{figure*}

\begin{equation}
H=\sum_{i=1}^N H(\theta_i,\theta_{i+1}),
\end{equation}
where 
$H(\theta,\psi)= -\frac{J}{2} \Big(\cos (\theta-\psi)-\cos (\theta+\psi)\Big) -\frac{h}{2} (\cos \theta+\cos \psi)$.

The key object to study the equilibrium properties is the transfer operator defined as
\begin{eqnarray}\label{eq:4}
	\mathsf{T} f(\theta)&\equiv&\int_{-\pi}^{\pi} \, d\theta \, e^{-\beta H(\theta,\psi)} f(\theta)
	\nonumber \\ 
	&\equiv& \int_{-\pi}^{\pi} \, d\theta  
	 T(\theta,\psi) f(\theta)\,,
\end{eqnarray}
where $T(\theta,\psi)=T(\psi,\theta)$ and $\beta=(k_B T)^{-1}$, with $k_B$ the Boltzmann constant.

This operator acts on square integrable functions and is such that the trace of its square is bounded, i.e., $\text{Tr } \mathsf{T}^2 \equiv \int d\theta d\psi \,\mathsf{T}^2(\theta,\psi) < \infty$. Associated with $\mathsf{T}$ there is an eigenvalue problem with the corresponding set of eigenvalues and eigenfunctions $\{\lambda_n,f_n(\theta)\}$, where $\lambda_n \ge \lambda_{n+1}$.
Knowledge of the first two leading eigenvalues of $\mathsf{T}$ is enough to characterize the main equilibrium properties. In particular, if $\lambda_0$ is the largest (in modulo) eigenvalue, then in the thermodynamic limit the partition function is given by
$
Z=\lambda_0^N$,
up to corrections of order $\mathcal{O}(e^{-N\log (\lambda_0/\lambda_1)})]$.

To determine the correlation length, we consider the two-point correlation function relative to the observables $s_1(\theta)$ and $s_2(\theta)$
\begin{equation}\label{eq:7}
C_{s_1,s_2}(l)\equiv \Big\langle s_2(\theta_i)s_1(\theta_{i+l})\Big\rangle,
\end{equation}
where the function $s(\theta)$ is associated with an operator $S$ and satisfies $S f(\theta)\equiv s(\theta) f(\theta)$. In terms of $T$ and $S$, the correlation function reads
\begin{equation}\label{eq:8}
C_{s_1,s_2}(l)=\lim_{N\to \infty} \frac{\text{Tr }S_1 T^l S_2 T^{N-l}}{\text{Tr } T^N} \quad \text{with } l\ge 0.
\end{equation} 
The decay of this function to its asymptotic value satisfies
\begin{equation}\label{key}
	C_{s_1,s_2}(l)-C_{s_1,s_2}(\infty)=\sum_{m\neq 0} c_m e^{-l/\xi_m},
\end{equation}
with $\xi_m^{-1}=\log{\frac{\lambda_0}{\lambda_m}}$. For large values of $l$ such decay is governed by the largest $\xi_m$, which is given by the correlation length $\xi$, defined as
\begin{equation}\label{eq:10}
	\xi^{-1}=\log \frac{\lambda_0}{\lambda_1}\,.
\end{equation}

To study the phase transition between the paramagnetic and the ferromagnetic phases, we introduce a single parameter $\epsilon=J-h$, such that $J=(1+\epsilon)/2$ and $h=(1-\epsilon)/2$. We analyze the behavior of the system in the range of values $\epsilon \in [-1,1]$.
The phase transition takes place as the temperature decreases and $\epsilon$ reaches a critical value
 \beqa
\epsilon^*=-\frac{1}{3}\rightarrow J=h+\frac{1}{3},
\eeqa
in stark contrast with the value in the TFIM $\epsilon^*=0$, with critical point at $J=h$. 

\begin{figure*}[t!]
\centering
\includegraphics[width=1\linewidth]{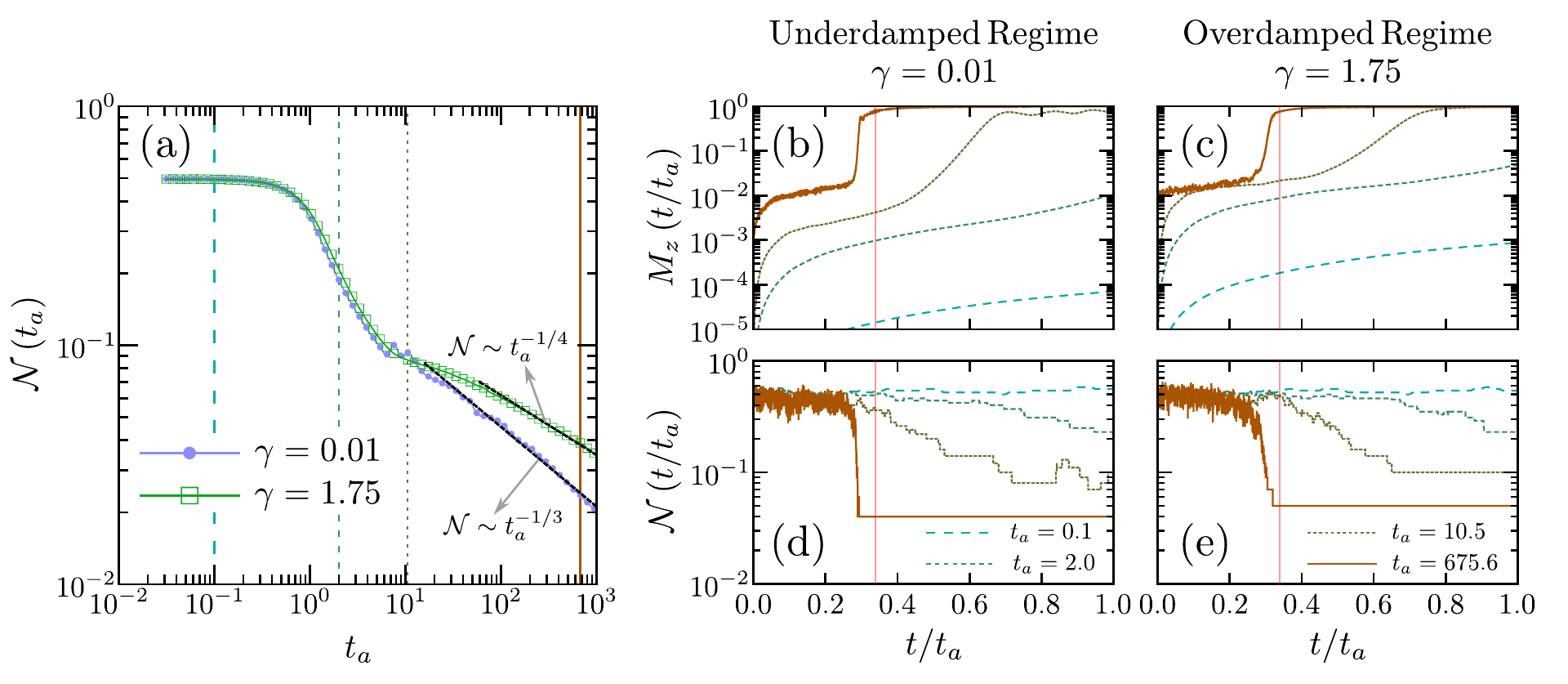}
\caption{\label{FigSR} Evolution of the number of topological defects and order parameter during the annealing protocol. 
(a) The average density of kinks as a function of the annealing time in the underdamped and overdamped limits. The number of kinks saturates to a plateau value at fast quenches. The subsequent dependence for slower ramps is unrelated to the KZM. Only in the limit of slow quenches is there a crossover to a power-law behavior that is described by the KZM. The numerical results match the theoretical KZM prediction in this regime. Four representative values of the annealing time $t_a$ are singled out (vertical lines) corresponding to the plateau, decay, crossover, and scaling regimes. For these values of the annealing time, the growth of the order parameter is shown in real-time during the course of the annealing schedule in the underdamped (b) and overdamped (c) regimes. The corresponding time evolution of the average kink number is shown in panels (d) and (e). The vertical red line in each panel represents the instant at which the critical point is crossed.
 }
\end{figure*}

\subsection{Equilibrium correlation length}

The phase transition occurs at the critical value $\epsilon^*$, around which the correlation length is expected to exhibit a universal power-law scaling. 
To explore it, we first focus on the kernel of the transfer operator $\mathsf{T}$, given by $\exp (-\beta H(\theta,\psi))$. Its major contribution in the vicinity of the maximum of the function $g(\theta,\psi)=-H(\theta,\psi)$ occurs for large values of $\beta$. An analysis of such function shows that its maximum values verify
\begin{equation}\label{eq:20}
	2\left(\frac{1+\epsilon}{1-\epsilon}\right) \sin \theta=\tan \theta\,.
\end{equation}
This equation has a single solution for $\epsilon<-1/3$ with  a single maximum located at $(\theta_0,\psi_0)=(0,0)$. When $\epsilon >-1/3$ the equation presents three possible solutions. In this case $(0,0)$ ceases to be a maximun and two new maxima appear. They start at $(0^{\pm},0^{\pm})$ when $\epsilon$ is infinitesimally greater than $-1/3$, and move continuously along the direction $\theta=\psi$ until reaching the value $(\theta_0,\psi_0)=\pm(\pi/2,\pi/2)$ when $\epsilon=1$. Close to the critical point $\epsilon^*=-1/3$ these two maxima are given by
\begin{equation}\label{eq:201}
	\theta_0=\psi_0=\pm 3 \sqrt{\frac{|\epsilon-\epsilon^{*}|}{2}}+\mathcal{O}(|\epsilon-\epsilon^{*}|^{3/2})\,.
\end{equation}
 
According to our previous discussion, at sufficiently low temperatures the kernel of the transfer operator can be approximated by its contribution in the vicinity of the maximum of the function $g$. More precisely, taking into account local contributions of $g$ up to second-order around the maximum, the kernel can be approximated by a Gaussian function. In this case, the eigenvalues and eigenfunctions can be analytically obtained. 
In the vicinity of a given maximum $(\theta_0,\psi_0)$, the function $g$ can be approximated by the second order expansion
	\begin{equation}\label{eq:18}
	\begin{split}
	g_0(\theta,\psi)=&g(\theta_0,\psi_0) +\frac{1}{2}\left[ g_{\theta \theta}(\theta_0,\psi_0)(\theta-\theta_0)^2\right.\\
	&+ 2 g_{\theta \psi}(\theta_0,\psi_0)(\theta-\theta_0)(\psi-\psi_0)\\
	&\left.+ g_{\psi \psi}(\theta_0,\psi_0)(\psi-\psi_0)^2 \right]+\cdots
	\end{split}
	\end{equation}

Then, the eigenvalue problem of $\mathsf{T}$ can be written in terms of the one corresponding to the operator $\mathsf{T}_0$, defined by
\begin{eqnarray}\label{eq:23}
	\mathsf{T}_0 f(\theta+\theta_0)&=&\int_{-\infty}^{\infty} \frac{d\theta}{\sqrt{\beta}} \, 
	e^{\frac{1}{2} (\theta,\psi)\cdot h(\theta_0,\psi_0)\cdot (\theta,\psi)^t} f(\theta+\theta_0) \nonumber \\ 
	&=&\lambda' f(\psi+\psi_0).
\end{eqnarray}
This operator has a Gaussian kernel with known eigenvalues and eigenfunctions, which lead to the following solutions to the problem (\ref{eq:23}):
\begin{eqnarray}\label{eq:24}
	\lambda_0=&& e^{\beta g(\theta_0,\psi_0)} \sqrt{\frac{2 \pi}{\beta}} \left({\sqrt{g^2_{\theta \theta}-g^2_{\theta \psi}}-g_{\theta \theta}}\right)^{-1/2},\nonumber\\ 
	\lambda_n=&&\lambda_0 e^{-n/\xi} \quad (n=1,2,\cdots),
\end{eqnarray}
with
\begin{eqnarray}
	\xi^{-1}= \log \left( 1+\frac{-g_{\theta \theta}-g_{\theta\psi}+\sqrt{g^2_{\theta \theta}-g^2_{\theta \psi}}}{g_{\theta\psi}}
	\right)\label{eq:25}.
\end{eqnarray}

The correlation length $\xi$ diverges at the transition point and encodes the transition at zero temperature. As shown in Fig. 1 in the main text, the sharp behavior around the critical point observed at very low temperatures is smoothed out as the temperature increases. 
The behavior of $\xi$ in the zero-temperature limit can be studied analytically. In the paramagnetic phase, $\epsilon<-1/3$, the correlation length in the vicinity of the critical point is set by
\begin{equation}\label{key}
	\xi \approx \frac{\sqrt{2}}{3}|\epsilon-\epsilon^*|^{-\frac{1}{2}} \left(1-\frac{9}{8}\sqrt{\frac{|\epsilon-\epsilon^*|}{2}}+\cdots\right),
\end{equation}
while in the ferromagnetic phase, $\epsilon>-1/3$, it reads
\begin{equation}\label{key}
	\xi \approx \frac{1}{3}|\epsilon-\epsilon^*|^{-\frac{1}{2}} \left(1-\frac{9}{16}|\epsilon-\epsilon^*|+\cdots \right).
\end{equation}
Our analysis thus reveals the critical exponent \beqa
\nu=1/2. 
\eeqa
This value further agrees with the one obtained from the numerical integration of the transfer matrix equation.

\section{Defect counting in real-time during the annealing schedule}\label{AppD}

The counting of kinks in a single realizations is performed by evaluating the kink number operator
\beqa
\mathcal{N}=\frac{1}{2}\sum_{i=1}^{N-1}\left[1-{\rm sgn}(\sin\theta_i){\rm sgn}(\sin\theta_{i+1})\right],
\eeqa
that upon acting on a given configuration of planar rotors ${\bm \theta}$ yields an integer kink number $n=0,1,2\dots$ as shown schematically in Fig.~\ref{FigScheme}. We also consider the order parameter which is given by the averaged absolute value of the local magnetization
\begin{equation}\label{op}
M_z(t)\,=\,\frac{1}{N}\sum_{i=1}^N\la|\sin\theta_i(t)|\ra,
\end{equation}
that reaches unit value in the ideal (anti)ferromagnetic configuration, i.e., when all spins are (anti)parallel to each other.
The growth of both of these quantities in real-time is presented in Fig.~\ref{FigSR} for various choices of the annealing time, spanning the different regimes in the dependence of the average kink density as a function of the annealing time.

\bibliography{Lang_SVMC_Bib}	
\newpage 
\clearpage
\widetext

\begin{center}
\textbf{\large ---Supplemental Material---\\
 Benchmarking quantum annealing dynamics: the spin-vector Langevin model}
 \end{center}
\setcounter{equation}{0}
\setcounter{figure}{0}
\setcounter{table}{0}
\setcounter{section}{0}
\setcounter{page}{1}
\makeatletter
\renewcommand{\theequation}{S\arabic{equation}}
\renewcommand{\thefigure}{S\arabic{figure}}
\renewcommand{\bibnumfmt}[1]{[S#1]}
\renewcommand{\citenumfont}[1]{S#1}

\begin{lstlisting}[language=Fortran]
ccccccccccccccccccccccccccccccccccccccccccccccccccccccccccccccccccccccccccccccccccccccc
cc
cc	Numerical integration of the stochastic SVL equations of motion.
cc
cc	Calculation of the number of kinks for a given value of the annealing time.
cc
cc	The GASDEV function of Numerical Recipes is used.
cc
cccccccccccccccccccccccccccccccccccccccccccccccccccccccccccccccccccccccccccccccccccccccc

	implicit none

	integer nr,nd                  ! nr: number of rotors
	parameter (nr=100,nd=2*nr) 
	
	integer ntime                  ! ntime: total number of time steps
	parameter (ntime=1200000)

	integer nta                    ! nta: number of stochastic trajectories
	parameter (nta=10000) 
	
	integer Jp(nr),Jm(nr),k,kk,ntan,itan,ita
	integer iseed,idum,i,ir,it         

	real*8 pi,kb,rm,temp,eps,gam,dd,bb,dt,dst
	real*8 h(nr),J(nr),th0(nr),p0(nr),t_an,tas
	real*8 tai,at_an(1:ntime),bt_an(1:ntime)
	real*8 time,at,bt,rn,dwi,rw(nd)
	real*8 va(nd),vb(nd),vgb(nd),vab(nd),y(nd),ynew(nd)
	real*8 ord_param,fi(nr),fip(nr),kinks,rnkink,r_kinks
	
	real*8 t_a
	
	real gasdev
	
	external gasdev,t_a

cc	Parameters

	pi=dacos(-1.d0)

	kb=1.d0             ! Boltzmann constant
cc
	rm=1.d0             ! mass of the rotors

	temp=1.d-3          ! temperature of the bath

	write (*,*) "(M,kb,T)",rm,kb,temp

	eps=0.05d0
	gam=eps*eps/(2.d0*kb*temp)      ! damping constant 

	dd=2.d0*rm*gam*kb*temp
	bb=dsqrt(dd)

	write (*,*) "eps,gam,bb",eps,gam,bb

ccc	The time step

	dt=1.d-3
	dst=dsqrt(dt)

cc	The connectivity function

	do ir=1,nr-1         ! i --> (i+1)
	Jp(ir)=ir+1
	end do
	            
	do ir=2,nr           ! i --> (i-1)
	Jm(ir)=ir-1
	end do

cc	The local magnetic field (h)

	do ir=1,nr
	h(ir)=0.d0
	end do

cc	The spin-spin coupling (J)

	do ir=1,nr
	J(ir)=1.d0
	end do

cc	The initial conditions (angles and momenta)

	do ir=1,nr
	th0(ir)=0.d0
	p0(ir)=0.d0   
	end do

cccc	THE TIME PROPAGATION (PLATEN - EXPLICIT SECOND ORDER)

ccc	The integration variable: y=(y(1),...,y(nr),y(nr+1),...,y(2*nr))
ccc	The angle coordinates: (y(1),...,y(nr))
ccc	The momentum coordinates: (y(nr+1),...,y(2*nr))

cc	The vector b with the temperatures

	call vector_b(nr,nd,bb,vb)

cc	File with a uniform distribution of integer values in the interval
cc	[1000,1000000]. These are the seeds used for the iteration in the 
ccc	in stochastic trajectories.

	open (unit=98,file="ran_data_seed.dat")

ccc	THE ANNEALING TIME

	k=63
	ntan=5427

	t_an=t_a(k)             ! this is the annealing time ta
	tas=dsqrt(t_an)

	write (*,*) "Anneling time",t_an

cc	Annealing schedule for 0 < t < ta (NASA DATA)

	open (unit=96,file="annealing_data_k63.dat")

	do itan=1,ntan
	read (96,*) tai,at_an(itan),bt_an(itan),kk
	end do

cc	The magnetic field is disconnect for t > ta

	do itan=ntan+1,ntime
	at_an(itan)=0.d0
	bt_an(itan)=bt_an(ntan)
	end do
cc	
	kinks=0.d0          ! number of kinks
	
cc	LOOP IN STOCHASTIC TRAJECTORIES	

	do ita=1,nta          

	read (98,*) iseed

	idum=iseed

ccc	The initial conditions for the numerical integration

	do ir=1,nr
	y(0*nr+ir)=th0(ir)
	y(1*nr+ir)=p0(ir)
	end do

c	LOOP IN TIME

	it=0

10	continue

	it=it+1

	time=dt*dfloat(it-1)

	at=at_an(it)                ! the A(t) function

	bt=bt_an(it)                ! the B(t)  function

cc	The vector with the noise. At each time step we pick a random value from 
cc	a Gaussian distribution, acting on the momentum coordinate of each rotor.
cc	The GASDEV function of Numerical Recipes is used. 

	do ir=1,nr
	rn=gasdev(idum)      ! random value from a Gaussian distribution    
	dwi=dst*rn           
	rw(0*nr+ir)=0.d0     ! there is not noise acting on the angle coordinates
	rw(1*nr+ir)=dwi      ! there is noise acting on the momentum coordinates
	end do

cc	Time integration scheme (according to Appendix B)

	call vector_a(nr,nd,gam,rm,Jp,Jm,J,h,at,bt,y,va)

	call vector_gb(nd,dt,rw,y,va,vb,vgb)

	call vector_a(nr,nd,gam,rm,Jp,Jm,J,h,at,bt,vgb,vab)

	do i=1,nd
	ynew(i)=y(i)+0.5d0*dt*(vab(i)+va(i))+rw(i)*vb(i)
	end do

	do i=1,nd
	y(i)=ynew(i)
	end do

cc	The order parameter

	ord_param=0.d0
	do ir=1,nr
	ord_param=ord_param+dabs(dsin(y(ir)))/dfloat(nr)
	end do

	if (ord_param.lt.0.9d0) goto 10

cc	The number of kinks

	do ir=1,nr-1
	fi(ir)=dsign(1.d0,dsin(y(ir)))
	fip(ir)=dsign(1.d0,dsin(y(ir+1)))
	end do

	rnkink=0.d0
	do ir=1,nr-1                   
	rnkink=rnkink+0.5d0*(1.d0-fi(ir)*fip(ir))
	end do
        
cc	The average over stochastic trajectories       
       
	kinks=kinks+rnkink/dfloat(nta)

	r_kinks=kinks/dfloat(nr)

	end do           ! LOOP IN STOCHASTIC TRAJECTORIES

cc	The density of kinks

	write (*,*) t_an,r_kinks,k

	stop

	end

cccccccccccccccccccccccccccccccccccccccccccccccccccccccccccccccccccccc

ccc	The annealing time

	real*8 function t_a(k)

	implicit none

	integer k

	real*8 pot

	pot=-5.d0+6.d0*0.02*(dfloat(k)-1.d0)

	t_a=2.d0**pot

	return

	end

cccccccccccccccccccccccccccccccccccccccccccccccccccccccccccccccccccccc

ccc	The vector b(t,y)

	subroutine vector_b(nr,nd,bb,vb)

	implicit none

	integer ib,nr,nd
	real*8 bb,vb(nd)

	do ib=1,nr
	vb(0*nr+ib)=0.d0
	vb(1*nr+ib)=bb
	end do

	return

	end

cccccccccccccccccccccccccccccccccccccccccccccccccccccccccccccccccccccc

ccc	The vector a(t,y)

	subroutine vector_a(nr,nd,gam,rm,Jp,Jm,J,h,at,bt,y,va)

	implicit none

	integer nr,nd,i,im,ip
	integer Jm(nr),Jp(nr)

	real*8 gam,rm,cpi,th_i
	real*8 th_im,th_ip
	real*8 at,bt,J(nr),h(nr),dh0,ss,dhp
	real*8 y(nd),va(nd)

cccc	i=2,nr-1 (internal rotors)

	do i=2,nr-1

	cpi=y(1*nr+i)               ! momentum
	th_i=y(i)                   ! angle (radians)

	im=Jm(i)                   ! (i-1)
	th_im=y(im)

	ip=Jp(i)                   ! (i+1)
	th_ip=y(ip)
cc
	dh0=at*dsin(th_i)

	ss=dsin(th_ip)+dsin(th_im)
	dhp=-bt*(J(i)*dcos(th_i)*ss-h(i)*dsin(th_i))
cc
	va(0*nr+i)=cpi/rm
	va(1*nr+i)=-0.5d0*(dh0+dhp)-gam*cpi

	end do

cccc	i=1 (first rotor)

	i=1

	cpi=y(1*nr+i)               ! momentum
	th_i=y(i)                   ! angle (radians)

	ip=Jp(i)                   ! (i+1)
	th_ip=y(ip)
cc
	dh0=at*dsin(th_i)

	ss=dsin(th_ip)
	dhp=-bt*(J(i)*dcos(th_i)*ss-h(i)*dsin(th_i))
cc
	va(0*nr+i)=cpi/rm
	va(1*nr+i)=-0.5d0*(dh0+dhp)-gam*cpi

cccc	i=nr (last rotor)

	i=nr

	cpi=y(1*nr+i)               ! momentum
	th_i=y(i)                   ! angle (radians)

	im=Jm(i)                   ! (i-1)
	th_im=y(im)

cc
	dh0=at*dsin(th_i)

	ss=dsin(th_im)
	dhp=-bt*(J(i)*dcos(th_i)*ss-h(i)*dsin(th_i))
cc
	va(0*nr+i)=cpi/rm
	va(1*nr+i)=-0.5d0*(dh0+dhp)-gam*cpi
cccc

	return

	end

cccccccccccccccccccccccccccccccccccccccccccccccccccccccccccccccccccc

ccc	The vector gam_bar = y + a dt + b dW

	subroutine vector_gb(nd,dt,rw,y,va,vb,vgb)

	implicit none

	integer l,nd

	real*8 dt,rw(nd),y(nd),va(nd),vb(nd),vgb(nd)

	do l=1,nd
	vgb(l)=y(l)+va(l)*dt+vb(l)*rw(l)
	end do

	return

	end

cccccccccccccccccccccccccccccccccccccccccccccccccccccccccccccccccccc

\end{lstlisting}
\end{document}